\documentclass[runningheads]{llncs}
\usepackage[T1]{fontenc}
\usepackage{graphicx}
\usepackage{hyperref}
\usepackage{color}

\usepackage{amssymb, amsmath,mathabx}
\usepackage{enumitem}
\setenumerate[0]{label=(\alph*)}
\usepackage{circledsteps}
\usepackage[mode=buildnew]{standalone}
\usepackage{tikz}

\usepackage{tikz}
\usetikzlibrary{arrows.meta,
								calc,
								positioning,
								shapes.misc}
\usepackage{fontawesome}
\usepackage{booktabs}

\tikzset{>={Latex[length=3mm]}}

\newcommand*{\ySpaceBetweenTrusteeAndEntities}{12em}

\tikzstyle{simple_entity} = [draw, rectangle,
minimum width = 2.8cm,
minimum height = 1.8cm,
thick
]

\tikzstyle{data_trustee} = [
draw,
rectangle,
minimum width = 3cm,
minimum height = 2.3cm, 
thick
]

\makeatletter
\tikzset{
    database top segment style/.style={draw},
    database middle segment style/.style={draw},
    database bottom segment style/.style={draw},
    database/.style={
        path picture={
            \path [database bottom segment style]
                (-\db@r,-0.5*\db@sh) 
                -- ++(0,-1*\db@sh) 
                arc [start angle=180, end angle=360,
                    x radius=\db@r, y radius=\db@ar*\db@r]
                -- ++(0,1*\db@sh)
                arc [start angle=360, end angle=180,
                    x radius=\db@r, y radius=\db@ar*\db@r];
            \path [database middle segment style]
                (-\db@r,0.5*\db@sh) 
                -- ++(0,-1*\db@sh) 
                arc [start angle=180, end angle=360,
                    x radius=\db@r, y radius=\db@ar*\db@r]
                -- ++(0,1*\db@sh)
                arc [start angle=360, end angle=180,
                    x radius=\db@r, y radius=\db@ar*\db@r];
            \path [database top segment style]
                (-\db@r,1.5*\db@sh) 
                -- ++(0,-1*\db@sh) 
                arc [start angle=180, end angle=360,
                    x radius=\db@r, y radius=\db@ar*\db@r]
                -- ++(0,1*\db@sh)
                arc [start angle=360, end angle=180,
                    x radius=\db@r, y radius=\db@ar*\db@r];
            \path [database top segment style]
                (0, 1.5*\db@sh) circle [x radius=\db@r, y radius=\db@ar*\db@r];
        },
        minimum width=2*\db@r + \pgflinewidth,
        minimum height=3*\db@sh + 2*\db@ar*\db@r + \pgflinewidth,
    },
    database segment height/.store in=\db@sh,
    database radius/.store in=\db@r,
    database aspect ratio/.store in=\db@ar,
    database segment height=0.1cm,
    database radius=0.25cm,
    database aspect ratio=0.35,
    database top segment/.style={
        database top segment style/.append style={#1}},
    database middle segment/.style={
        database middle segment style/.append style={#1}},
    database bottom segment/.style={
        database bottom segment style/.append style={#1}}
}
\makeatother

\tikzstyle{mydatabase} = [database,
database segment height=0.5em, database radius = 1em,
]

\tikzstyle{steward_database} = [mydatabase,
database top segment = {fill=white},
database middle segment = {fill=white},
database bottom segment = {fill=white}
]

\tikzstyle{exchange_database} = [mydatabase]
\newcommand{\mytable}[5]{
		\node[below= of #1 , anchor = north west] (T_#3_left) {\small
			\mytablee{#3}{1}{#4}{#5}
	};
		\node[below= of #2, anchor= north east] (T_#3_right) {\small
			\mytablee{#3}{m_#3}{{#4'}}{{#5'}}
		};
		\node (T_#3_dots) at ($(T_#3_left.east)!.5!(T_#3_right.west)$) {\kern0.48em\ldots};
		
}

\newcommand{\mytablee}[4]{
			\begin{tabular}{ |c | c| }
				$\mathbf{T_#1[#2]}$ & $\mathbb{A}(T_#1[#2])$ \\
				\midrule
				$uid_#3$ & $T_#1[1,#2]$  \\
				$uid_#4$ & $T_#1[2,#2]$ \\[-1ex]
				$\vdots$ & $\vdots$  \\
			\end{tabular}
}

\usepackage[english,linesnumbered,ruled,vlined, noend]{algorithm2e}

\usepackage{todonotes}
\usepackage{csquotes}

\usepackage{framed}
\usepackage[amsmath,framed,hyperref]{ntheorem}
\theoremstyle{break}
\newframedtheorem{Problem}{Problem}
\theoremstyle{plain}
\theoremheaderfont{\itshape}
\theorembodyfont{\normalfont}
\theoremseparator{.}
\newtheorem{statement}{Statement}

\makeatletter
\@addtoreset{statement}{subsection}
\makeatother

\usepackage{cleveref}
\crefname{Problem}{problem}{problems}
\crefname{statement}{statement}{statements}
\Crefname{equation}{Eq.}{Eqs.}

\begin{document}
%
\title{PrivTru: A Privacy-by-Design Data Trustee\\Minimizing Information Leakage}
%
%
\author{Lukas~Gehring\inst{1}\orcidID{0009-0006-7876-861X} \and%
Florian~Tschorsch\inst{1}\orcidID{0000-0001-6716-7225}}
\authorrunning{L. Gehring, F. Tschorsch}
%
\institute{Technische Universität Dresden, 01062 Dresden, Germany
\email{\{lukas.gehring,florian.tschorsch\}@tu-dresden.de}}
\maketitle              
\begin{abstract}
Data trustees serve as intermediaries
that facilitate secure data sharing between independent parties.
This paper offers a technical perspective on data trustees,
guided by privacy-by-design principles.
We introduce PrivTru, an instantiation of a data trustee
that provably achieves optimal privacy properties.
Therefore, PrivTru calculates the minimal amount of information
the data trustee needs to request from data sources to respond to a given query.
Our analysis shows that PrivTru minimizes information leakage
to the data trustee, regardless of the trustee's prior knowledge,
while preserving the utility of the data.

\keywords{Data Trustee \and Privacy by design \and Privacy Engineering}
\end{abstract}

\section{Introduction}\label{intro}
The concept of a data trustee (or sometimes data trust)
facilitates a data economy that aligns
with privacy goals and data protection regulations, particularly in
the EU~\cite{reiberg_datentreuhander_2023,blankertz_what_2021,aline_blankertz_designing_2020}.
It addresses challenges in sharing data between
independent data sources and data receivers.
This is particularly
relevant in fields such as medicine, public administration,
mobility, and other domains where data can drive innovation and
create value. Politically, data trustees are presented as an alternative
to the dominance of data platforms, which risk creating a \enquote{data
oligopoly} that centralizes power and limits
competition~\cite{reiberg_datentreuhander_2023}.

Most discussions have focused on the organizational aspects~\cite{blankertz_what_2021},
situating data trustees as a third party between data sources
(entities who \emph{have} data)
and data receivers (entities who \emph{want} data).
A core idea of the concept is that sources and receivers are independent systems,
operated by any entity, enabling data sharing (and potentially linking)
across institutional boundaries.
To fulfill this role, a data trustee needs to be constituted as an independent organization,
which has no interest in the data itself.
As such it can reasonably balance the interests of the involved parties.
%
The primary task of a data trustee is, thus, to manage data 
\enquote{on behalf of and in the interest of}~\cite{reiberg_datentreuhander_2023}
the data sources.

The literature~\cite{aline_blankertz_designing_2020,reiberg_datentreuhander_2023} categorizes trustees into two types:
\emph{data stewards} and \emph{data exchange}.
In the former, the trustee stores the data,
whereas in the latter, the trustee facilitates matching
between sources and receivers and relays the data without storing it.

In this paper, we present a technical perspective on data trustees,
focusing specifically on data stewards and data exchanges.
This work addresses a gap in the discussion on data trustees,
which has largely been explored from organizational, political, and legal perspectives.
By applying Hoepman's privacy design strategies~\cite{hoepman_privacy_2014},
we identify the challenges of implementing privacy-respecting data trustees
and demonstrate that the \emph{data exchange} model is preferable
from a privacy-centric standpoint.

We then introduce \emph{PrivTru}, a data exchange designed to be private by design.
To this end, we extend relational algebra,
enabling the formulation of individual subqueries for generalized data
from the data sources.
By leveraging this approach, PrivTru outsources pre-processing tasks to the data sources,
ensuring that only the necessary information is retrieved to answer a given query.
We further prove that our instantiation is optimal in minimizing the information
the data trustee learns about the provided data.
The proposed solution achieves this without any loss of utility
and remains optimal regardless of the prior knowledge
the central exchange may have about the input data.
Accordingly, our contributions can be summarized as follows:
\begin{itemize}
\item We formulate a technical perspective on data stewards and data exchanges,
  the two primary types of data trustees (\Cref{model}).
\item We use Hoepman's privacy design strategies~\cite{hoepman_privacy_2014}
  to analyze the privacy properties of the proposed models (\Cref{pa}).
\item We introduce \emph{PrivTru}, an instantiation of a data exchange
  aligned with privacy engineering principles (\Cref{instantiation}).
\item We analyze the information gain of the central data exchange in \emph{PrivTru}
  and prove that it is minimal compared to all other
  relational data exchange implementations (\Cref{analysis}).
\end{itemize}

\section{Technical Perspective on Data Trustees}\label{model}

In this section, we provide a technical perspective
on the notion of a \emph{data trustee}
by introducing two models capturing its main variations:
\emph{data steward} and \emph{data exchange}.
To encompass most concepts covered by this notion,
the models focus on the core functionality of a data trustee,
which is providing access to data from \emph{data
sources} to \emph{data receivers}.
Apparently, given the broad understanding of a data trustee,
there may be some instantiations of data trustees, stewards, and exchanges
that are not compatible with our model.
We additionally consider the question of bootstrapping the connection
between sources, receivers, and the trustees,
notably the implementation of access control, to be out of scope.
Nevertheless, the models are designed to be extensible,
allowing for adaptation to specific organizational and legal requirements
of data trustees or to provide better guidance for practical implementation.

\par
In the data trustee model, the third party sitting between the
sources and the receivers is a \emph{steward} or an \emph{exchange}.
These are differentiated by their mandate to store the processed data,
which leads to distinct data flows when in operation.
A \emph{steward} stores data on behalf of the sources,
while an \emph{exchange} merely processes the information from the sources,
without storing any data.
\Cref{fig_ds} illustrates the architecture and functionality of a data steward.
In the setup phase, all sources send \emph{all} of their data
to the steward~(Step~\Circled{0}), which then stores it centrally.
Subsequently, the steward provides one or more API~endpoints for the receivers
to query the data~(Step~\Circled{1}).
It processes these queries by sending results back
to the corresponding receiver~(Step~\Circled{2}).

The architecture of a data exchange is depicted in \Cref{fig_de}.
During the bootstrapping process~(Step~\Circled{0}),
the central entity only receives the schema of the data provided by the sources.
Exchanges provide the same interface as stewards,
enabling receivers to query information
and obtain results~(Steps~\Circled{1} and \Circled{4}).
However, unlike stewards, exchanges do not store any data.
Instead, upon receiving a query~(Step~\Circled{1}),
they generate subqueries for each data source and assembles the subresults
to answer the full query~(Steps~\Circled{2} and~\Circled{3}).

\begin{figure}[t]
\centering
\begin{tikzpicture}
	
%

	\node[
	data_trustee
	] (data_steward) at (0,0){};
%
	\node (data_steward_heading) at ([yshift=-0.3cm] data_steward.north) {\textbf{Steward}};


		\node[steward_database,xshift=2mm, yshift=1.5mm] at (data_steward) (db_3)  {};
		\node[steward_database,yshift=-0.5mm] at (data_steward) (db_2)  {};
		\node[steward_database,xshift=-2mm, yshift=-2.5mm] at (data_steward) (db_1)  {};
		\node[yshift=3mm] at (data_steward.south) {\textbf{Data}};

	\coordinate[left = \ySpaceBetweenTrusteeAndEntities of data_steward] (cor_data_source_origin);
	\node[simple_entity,draw=gray, xshift=-2.5mm, yshift=2.5mm] at (cor_data_source_origin) (n_rect) {};
	\node[simple_entity,draw=gray, fill=white,
	] at (cor_data_source_origin) (2_rect) {};
	\node[simple_entity, 
	xshift=2.5mm, yshift=-2.5mm,
	fill=white] at (cor_data_source_origin) (1_rect) {Sources};

	\coordinate[right = \ySpaceBetweenTrusteeAndEntities of data_steward] (cor_data_reciver_origin);
	\node[simple_entity,draw=gray,
	xshift=2.5mm,
	yshift=2.5mm]
	at (cor_data_reciver_origin) (data_processor_p) {};
	\node[simple_entity,fill=white,
	draw=gray,
	] at (cor_data_reciver_origin) (data_processor_2) {};
	\node[simple_entity, fill=white,
	xshift=-2.5mm,
	yshift=-2.5mm] at (cor_data_reciver_origin) (data_processor_1) {Recievers};

	\draw[->,thick,transform canvas={yshift=-2.5mm}]
	(1_rect.east |- 2_rect) -- (data_steward) node[midway, sloped, above] {\Circled{0}~Data};

	\draw[->,thick,transform canvas={yshift=2.5mm}] (data_processor_1.west |- data_processor_2) -- (data_steward) node[midway,above] {\Circled{1}~Queries};
	\draw[<-,thick,transform canvas={yshift=-7.5mm}] (data_processor_1.west |- data_processor_2) -- (data_steward) node[midway,above] {\Circled{2}~Results};

\end{tikzpicture}
\caption{Illustration of a data steward.}
\label{fig_ds}
\end{figure}
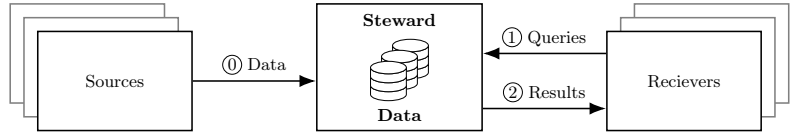
\begin{figure}[t]
\centering
\begin{tikzpicture}

	\node[
	data_trustee
	] at (0,0) (data_exchange) {\huge\faExchange};
	\node (data_exchange_heading) at ([yshift=-0.3cm] data_exchange.north) {\textbf{Exchange}};

	\coordinate[left = \ySpaceBetweenTrusteeAndEntities of data_exchange] (cor_data_source_origin);
	\node[simple_entity,draw=gray, xshift=-2.5mm, yshift=2.5mm] at (cor_data_source_origin) (n_rect) {};
	\node[simple_entity,draw=gray, fill=white] at (cor_data_source_origin) (2_rect) {};
	\node[simple_entity, fill=white,
	xshift=2.5mm, yshift=-2.5mm
	]
	at (cor_data_source_origin) (1_rect) {};
	\node[exchange_database,
	label=above:Sources,
	label=below:\textbf{Data}]
	at ([yshift=-2.5mm,xshift=2.5mm]cor_data_source_origin) (db_2)  {};

	\coordinate[right = \ySpaceBetweenTrusteeAndEntities of data_exchange] (cor_data_reciver_origin);
	\node[simple_entity,draw=gray, xshift=2.5mm, yshift=2.5mm] at (cor_data_reciver_origin) (data_processor_p) {};
	\node[simple_entity,fill=white, draw=gray] at (cor_data_reciver_origin) (data_processor_2) {};
	\node[simple_entity, fill=white,
	xshift=-2.5mm,
	yshift=-2.5mm
	] at (cor_data_reciver_origin) (data_processor_1) {Recievers};

	\draw[->,thick,transform canvas={yshift=5mm}] (1_rect.east |- 2_rect) -- (data_exchange) node[midway,above] {\Circled{0}~Schemas};
	\draw[<-,thick,transform canvas={yshift=-2.5mm}] (1_rect.east |- 2_rect) -- (data_exchange) node[midway,above] {\Circled{2}~Subqueries};
	\draw[->,thick,transform canvas={yshift=-10mm}] (1_rect.east |- 2_rect) -- (data_exchange) node[midway,above] {\Circled{3}~Subresults};

	\draw[->,thick,transform canvas={yshift=2.5mm}] (data_processor_1.west |- data_processor_2) -- (data_exchange) node[midway,above] {\Circled{1}~Queries};
	\draw[<-,thick,transform canvas={yshift=-7.5mm}] (data_processor_1.west |- data_processor_2) -- (data_exchange) node[midway,above] {\Circled{4}~Results};

\end{tikzpicture}
\caption{Illustration of a data exchange.}
\label{fig_de}
\end{figure}
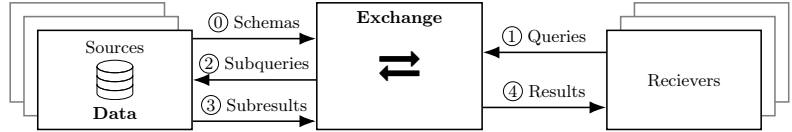

\section{Privacy Analysis of Data Trustees}\label{pa}

A system that prioritizes privacy should be designed with respect to
\emph{privacy-by-design} strategies.
In this section, we apply these design strategies to our two data trustees models
and argue that \emph{data exchanges} are preferable to \emph{data stewards}
under these principles.

\subsubsection{Privacy Design Strategies}\label{subsec_pg}

To bridge the gap between privacy requirements for IT systems
driven by ethical and legal requirements and software development,
Hoepman proposed eight privacy design strategies~\cite{hoepman_privacy_2014}.
According to this perspective,
a system's design plays a significant role in shaping its level of privacy.
Therefore, privacy considerations should be taken into account from the
very beginning, when the foundational concepts and ideas for a new
system are first being developed, which includes our models of data
trustees.
His contributions are categorized into data-oriented strategies
and process-oriented strategies.
Since we focus on the technical side of data trustees,
our analysis is limited to the former.
Of course, process-oriented strategies must also be considered
when deploying a data trustee.
The data-oriented privacy strategies as formulated by
Hoepman~\cite{hoepman_privacy_2014} are as follows:

\begin{itemize}
\item \textsc{Minimize}: The amount of personal data that is processed
  should be limited to the minimal amount possible.

  \item \textsc{Hide}: Any personal data, and their relationships,
    should be concealed from plain view.

  \item \textsc{Separate}: Personal data should be processed in a
    distributed manner, using separate compartments whenever possible.

  \item \textsc{Aggregate}: Personal data should be processed at the
    highest level of aggregation and with the least detail necessary
    for it to remain useful.
\end{itemize}

\subsubsection{Comparing Data Stewards and Data Exchanges}

In the following analysis, we compare the data trustee models (as
outlined in \Cref{model}) based on the data-oriented privacy strategies.
Hereby we focus on the differences between the two models and
always assume the same premises when possible.

{\textsc{Minimize}}: Data trustees generally allocate information
from multiple data sources to be jointly provided to one or more data
receivers. This is a problem for data minimization because when two or
more personal data points concerning one individual are joined, the
generated information is in most cases more privacy-sensitive than the
two data points alone since it increases the risk of
\emph{re-identification attacks}~\cite{samarati_protecting_1998}.
Still, one
can argue that the data is sufficiently minimized if two requirements
are fulfilled. Firstly, the data sources need to make sure, that the
data they provide is already minimized. Secondly, the purpose of the
whole system is formulated broadly. So that it encompasses all use cases
where any data might be requested. If the
sources give their consent to the possible wide range of data usage,
such usage is not a privacy violation.
If a data trustee satisfies these conditions, we
can consider the system to be in line with \textsc{Minimize}.
In the data exchange model, data sources can minimize queries locally,
thereby enhancing privacy when compared to data stewards.

\textsc{{Hide}}:
For data trustees, it is essential to ensure
that data remains confidential.
Encrypting connections secures the content
during transfer.
However, when data requests involve only a subset of sources,
it may also be necessary to obscure traffic patterns---%
especially when a single data source corresponds to an individual
or a small group, a challenge that solely affects data exchanges.
In such cases, obfuscation techniques can mitigate risks.
This leaves data exchanges with a slight but manageable disadvantage
compared to data stewards.

\textsc{Separate}: Organizational data trustees are formulated
as centralized systems, which opposes the \textsc{Separate} design
strategy.
As discussed in \Cref{model}, a centralized organizational concept
does not necessarily demand a centralized technical model.
The data exchange model is fundamentally embracing the \textsc{Separate} strategy,
while the data steward model is breaking it.
This is especially true if a data exchange is instantiated by a distributed protocol between the data source and receivers, and not through a centralized service.
Regarding \textsc{Separate}, data exchanges are to be favored over data stewards.

\textsc{Aggregate}:
Data aggregation can occur at the data sources or by the data steward/exchange.
For certain techniques, such as k-anonymity~\cite{samarati_protecting_1998},
access to all data points is critical for a reasonable privacy-utility tradeoff.
Thus, aggregation strategies must be carefully tailored to each use case.
Data exchanges are generally preferred due to their flexibility
in determining aggregation levels per query,
whereas the data steward model requires defining these levels
during the bootstrapping phase.

\subsubsection{Summary of Comparison}
Summarizing the analysis, we observe that data exchanges are mostly
preferable from a privacy perspective. This is largely due to the
\textsc{Separate} strategy, which data exchanges fully implement and
data stewards compromise by default. For \textsc{Hide}, both models
are comparable, although data exchanges are more prone
to information leakage via side channels.
\textsc{Minimize} \& \textsc{Aggregate} can be achieved by both models;
however, the data exchange model conceptually offers more flexibility,
as the data sources can minimize or aggregate the data on a per-query basis.

\section{PrivTru: Minimizing Information Leakage}\label{instantiation}
In this section, we introduce \emph{PrivTru},
a data trustee designed with \emph{privacy-by-design} principles.
Our analysis revealed that designing PrivTru as a data exchange
is preferable for aligning with privacy principles.
Yet, implementing \textsc{Minimize} and \textsc{Aggregate} poses challenges.
In our work, we focus on \textsc{Minimize}
and argue that \textsc{Aggregate} can be addressed similarly,
as discussed in \Cref{conclusion}.
We specifically examine how queries to PrivTru are structured
and how subqueries are derived.
%


\subsubsection{Data and Query Format}\label{data_and_query_format}
For PrivTru, we adopt the commonly used relational database model,
where data is stored in tables~$T$,
also referred to as \emph{relations}.
Each table consists of multiple columns, identified by a set of
\emph{attributes}~$\mathbb{A}(T) = \{A_1, \dots A_n \}$, with $n > 0$.
Each attribute~$A_i$ is associated with a \emph{domain}~$\mathbb{D}(A_i)$,
which defines the set of permitted values in this column.
Accordingly, the datapoints~$t$ in~$T$ are the rows of the table.
Formally, they are tuples of length~$n$
with~$t_i \in \mathbb{D}(A_i)$ for~$i = 1, \dots n$.
In the system model of data exchanges,
there are multiple data sources~$S_1, \dots S_s$,
each providing a portion of the data.
For readability, we assume that each source~$S_i$ contributes
exactly one table~$T_i$ to the system.
In addition, the attributes across all tables
are assumed to be pairwise distinct,
meaning that every attribute in the system is present
in exactly one table.
Note, that data trustees in general
and data exchanges specifically
are not tied to a specific data model.

Using relational databases as our data model
naturally leads to relational algebra
as the foundation for our query model.
Specifically, we adopt the relational algebra
as introduced by Codd~\cite{codd_relational_1970}.

\begin{definition}[Relational Algebra~\cite{abiteboul_foundations_1995}]
	The \emph{relational algebra} consists of three primitive operators,
	working on tables (or relations) \(T\).
\begin{enumerate}
\item
  \emph{Selection}: \(\sigma_{F}(T)\) selects tuples in~\(T\)
  that fulfill a given propositional formula~\(F\). The formula~\(F\)
  consists of the logical operators~\(\land , \vee ,\neg\),
  connecting literals of the form \texttt{true}, \texttt{false},
	\(A\theta B\), and \(A\theta c\)
	with $A,B \in \mathbb{A}(T), c \in \mathbb{D}(A)$, and
	$\theta \in \{ = ,<\}$.
\item
\emph{Projection}: \(\pi_{\left( A_{1},\ldots A_{n} \right)} (T)\)
  outputs a new relation containing the columns identified by
  \(A_{1},\ldots A_{n}\) in \(T\).
\item
	\emph{Join}: (T \(\Join\) S) outputs a relation $O$ with $\mathbb{A}(O) = \mathbb{A}(T) \cup \mathbb{A}(S)$.
	Where every~$o$ over $\mathbb{A}(O)$ is in $O$ if and only if its restrictions are in the original table, that is, $o[\mathbb{A}(T)] \in T$ and $o[\mathbb{A}(S)] \in S$.
	In case that $\mathbb{A}(T) \cap \mathbb{A}(S) = \emptyset$, the join operator is equal to calculating the cross-product.
\end{enumerate}
\end{definition}

We now present an extension of the relational algebra,
which we will use in the next subsection to achieve \textsc{Minimize} in PrivTru.
This extension enables data sources to evaluate basic propositions on their local entries
and to communicate the corresponding truth-values without revealing the underlying data.
To enhance readability, we
write $(\neg)$, to represent optional negations.
As a result, the following $(\neg)$ are also evaluated as negation in the given context.

\begin{definition}[Relational Algebra with Propositional Projections]\label{def:rapp}
The \emph{relational algebra with propositional projections~(RAPP)} consists of
the three primitive operators of the relational algebra, where the \emph{projection} operator is replaced by the following:
\(\pi_{\left( A_{1},\ldots A_{n} \right)} (T)\)
  outputs a new relation containing the columns identified by
  \(A_{1},\dots A_{n}\) in \(T\). Additionally, any
	\(A_{i}\) can have the form \(p: \bigvee_{i}(\neg)\gamma_i\),
	where $p$ is any string and
  all \(\gamma_i\) are literals of the form \texttt{true}, \texttt{false},
	\(A\theta B\) or \(A\theta c\), with $A,B \in \mathbb{A}(T), c \in \mathbb{D}(A)$ and
	$\theta \in \{ = ,<\}$.
	In this case, the resulting relation
\(R = \pi_{\left( A_{1},\ldots A_{n} \right)} (T)\) consists of a new
  column identified by~$p$, where each tuple in~$R$
  holds \texttt{true} or \texttt{false} depending on the evaluation
  of~\(\bigvee_{i}(\neg)\gamma_i\) on that tuple in~$T$.
\end{definition}

For example, \Cref{tab_T} shows a table of persons
with their names, ages, and income.
Using the propositional projections of RAPP,
we can query this data to produce a table that indicates only
whether each person is older than 30~(see~\Cref{tab_T_over_30}).
This projection is also supported in modern SQL dialects.

\begin{table}
\setlength{\tabcolsep}{12pt}
\parbox{.45\linewidth}{
\centering
\caption{$T$}\label{tab_T}
\begin{tabular}{ccc}
	\toprule
	\textbf{Name} & \textbf{Age} & \textbf{Income}\\
		\midrule
		Alice & \(30\) & 44k\\
		Bob & \(33\) & 40k\\
		Carol & \(50\) & 52k\\
		Eve & \(42\) & 66k\\
		\bottomrule
\end{tabular}
}
\hfill
\parbox{.45\linewidth}{
\centering
\caption{$\pi_{\text{Name, over\_30: age} > 30} T$}\label{tab_T_over_30}
\begin{tabular}{cc}
		\toprule
		\textbf{Name} & \textbf{over\_30}\\
		\midrule
		Alice & \texttt{false} \\
		Bob & \texttt{true} \\
		Carol & \texttt{true} \\
		Eve & \texttt{true}\\
		\bottomrule
\end{tabular}
}
\end{table}

\subsubsection{Calculating Subqueries for Relational Data Exchanges}\label{problem}

Every data exchange compatible with our model
must have a strategy to derive subqueries from the main query.
We formally define this problem as follows.

\begin{Problem}[CalculateSubqueries (CS-Problem)]\label{problemOne}
	\emph{Input}: Query $q$ in the language of relational algebra
	over relations $T_1, \dots T_s$.\\
	\emph{Output}: For $i = 1, \dots s$, subqueries $\widehat{q_i}$
	in the language of RAPP over
	$T_i$ and a collecting query $\widehat{q}$ over the relations $R_1, \dots R_s$.
	So that for every set of tables $T_1, \dots T_s$, the evaluation of $q$ on $T_1 \dots T_s$ is equal to the evaluation of~$\widehat{q}$ on $R_1, \dots R_s$, where $R_i$ is the result of the evaluation of $\widehat{q_i}$ on $T_i$.
\end{Problem}

Every algorithm~$\mathcal{A}$ that solves \Cref{problemOne}
can be canonically transferred into a data exchange,
given we have algorithms for evaluating queries in relational algebra and RAPP.
For this, the exchange calculates~$\widehat{q_1}, \dots \widehat{q_s}$
from the input query~$q$ (using~$\mathcal{A}$)
and sends these subqueries to the respective data sources $S_1, \dots S_s$.
Each source evaluates its query on its local data
and returns the results~$R_1, \dots R_s$ to the exchange.
The exchange then evaluates~$\widehat{q}$ over $R_1, \dots R_s$
and sends the result back to the data receiver.

A trivial algorithm to solve \Cref{problemOne} is to
set~$\widehat{q_i} := T_i$ and set $\widehat{q}$ to~$q$.
In this case, the exchange requests all data from all sources,
enabling it to calculate $q$ locally.
However, this solution directly contradicts the privacy analysis in \Cref{pa}.
It violates \textsc{Minimize} and undermines the spirit of \textsc{Separate},
since the exchange gains complete knowledge of the data on any query.

To develop a more privacy-friendly solution,
we first need to introduce some theoretical preliminaries.
\Cref{def:enf} defines a form for expressing queries in relational algebras,
based on the commonly used normal form~\cite{abiteboul_foundations_1995}.
\Cref{lemma_rewrite} then presents rewrite rules for these queries.

\begin{definition}[Extended Normal Form]\label{def:enf}
A query~\(q\) in the language relational algebra is in \emph{extended normal form} if:
\begin{enumerate}
\item
  \(q\) is in normal form, which means \(q\) can be written as $
		q = \pi_{\beta} 
	\sigma_{F}
		(
			 T_{1}
			\Join \ldots \Join
			 T_{k}
		)
		$.
\item
  \(F\) is in conjunctive normal form, meaning it can be written
  as
  $F := \bigwedge_{i}x_{i} := \bigwedge_{i}\bigvee_{j}(\neg)\gamma_{i,j}$,
  where \(x_{i}\) are clauses that are disjunctions of the literals
  \(\gamma_{i,j}\).
\item
	If $F$ is not a trivial tautology (F = \texttt{true})
	or trivially not satisfiable (F = \texttt{false}), then
	$F$ is satisfiable and not a tautology and
	every clause  $x$ in $F$ is satisfiable and not a tautology and
	every literal $\gamma$ in $F$ is satisfiable and not a tautology.
\end{enumerate}
\end{definition}
%
Database theory established that every query~$q$ can be rewritten in normal form,
and every propositional formula~$F$ can be rewritten in conjunctive normal form~\cite{abiteboul_foundations_1995}.
Furthermore, \Cref{def:enf}(c) can be achieved using rewrite rules
and an algorithm that checks for satisfiability.
Consequently, every query~$q$ can be expressed in extended normal form.
\Cref{def:enf} makes the queries manageable in our system.
\Cref{def:enf}(c) is particularly useful,
as it enables mapping literals to the tables that determine their values.
More precisely,
for any set of literals~$\Gamma$ used in a RAPP query,
let $\mathbb{T}(\Gamma)$ be the set of tables~$T$
containing $A \in \mathbb{A}(T)$,
which is used in a literal $\gamma \in \Gamma$.
We sometimes omit parentheses when using~$\mathbb{T}$.
Since we assume that attributes are unique across all tables
and because of \Cref{def:rapp,def:enf}(c),
$1 \leq \mathbb{T}(\gamma) \leq 2$ holds for all $\gamma$ in our scenario.

The following lemma provides useful rewrite rules for propositions and queries, which we will use to build an algorithm that solves~\Cref{problemOne}.

\begin{lemma}\label{lemma_rewrite}
	For every set of relations $R_1, \dots R_s$,
	proposition formula $F$,
	literals~$\gamma, \gamma_1, \dots \gamma_m$,
	list of attributes $\beta$
	and $i \in \{1, \dots s\}$:
	\begin{enumerate}
		\item
			If $F := F' \wedge (x)$, where $x = \bigvee_{i=1}^m (\neg_i)\gamma_i$ and $\mathbb{T}(\gamma_1,\dots \gamma_m) = \{R_i\}$, then
			$
				\sigma_{F}
				(
					R_1 \Join \cdots \Join R_n
					)
					\equiv
				\sigma_{F'}
					(
						R_1 \Join \cdots
						\sigma_{x} (R_i)
						\cdots \Join R_n
				)
				$.
		\item
			If $F := F' \wedge (\neg_1)\gamma_1 \vee \dots \vee (\neg_m)\gamma_m$ and for a subset $\Gamma \subset\{\gamma_1,\dots \gamma_m\}, \mathbb{T}(\Gamma) = \{R_i\}$, then 
			$
				\sigma_{F}
				(
					R_1 \Join \cdots \Join R_n
					)
					\equiv
			\sigma_{F' \wedge
				(\bigvee_{\gamma_i \in \{\gamma_1 ,\dots \gamma_m\} \setminus \Gamma} (\neg_i)(\gamma_i)
				\vee p = \texttt{true})}
				(
					R_1 \Join \cdots
					\pi_{p: \bigvee_{\gamma_j \in \Gamma}(\neg_j)(\gamma_j)}R_i
					\cdots \Join
					R_n
					)
			$.
		\item
			Let $\mathbb{A}(R_i, F, \beta)$ denote the set of attributes $A$ in $\beta$ and $F$, with $A \in \mathbb{A}(R_i)$, then
			$
			\pi_{\beta}\sigma_{F}(R_1 \Join \cdots \Join R_n)
			\equiv
			\pi_{\beta}\sigma_{F}(R_1 \Join \cdots
			\pi_{\mathbb{A}(R_i, F, \beta)}(R_i)
			\cdots \Join R_n)
			$.
		%
	\end{enumerate}
\end{lemma}
We can now introduce \Cref{algo_distributed_queries},
which is a solution to \Cref{problemOne} as we will prove with \Cref{theoremOne}.
In the algorithm, we treat attribute lists and propositional formulas as sets.
More precisely, for a projection $\pi_{\beta}$ in RAPP,
$\beta$ is a set of attributes and constructs of the form $p: \gamma$.
Propositional formulas~$F$ (in conjunctive normal form) are viewed as a set of sets
containing literals~$\gamma$, where each inner set corresponds to a clause $x$.
The conjunction~$F$ is formed by combining these clauses,
while each clause represents a disjunction of its literals.

\begin{algorithm}
	\KwData{Query $q = \pi_{\beta}\sigma_{F}(T_{1} \Join \ldots \Join T_{s})$
	in extended normal form with $F := \bigwedge_{k=0}^{n}x_k := \bigwedge_{k=0}^{n}\bigvee_{j=0}^{n_k}(\neg)\gamma_{k,j}$.}
	\If{$F = \texttt{false}$}{
		\Return{$q^\varnothing, \dots q^{\varnothing}$}\tcp*[l]{$s+1$ times the empty query $q^{\varnothing}$}
	}
  \(\widehat{q} \leftarrow \varnothing\)\;
  \(\widehat{F} \leftarrow F\)\;
	\ForEach{$i \in \{1, \dots s\}$} {
    \(\widehat{q_{i}} \leftarrow \varnothing\);
    \(\beta_{i} \leftarrow \varnothing\);
    \(F_{i} \leftarrow \varnothing\);
    \(\Gamma_{i} \leftarrow \varnothing\);
	}
	\ForEach(\tcp*[h]{$x$ is a set representing a clause}){$x \in F$}{\label{line_foreach_1}
		\If(\tcp*[h]{Use \Cref{lemma_rewrite}(a)}){$|\mathbb{T}(x)| = 1$}{\label{line_lemma_a_start}
			\ForEach{$T_i \in \{T_1, \dots T_s\}$}{
				\If{$\mathbb{T}(x) = \{T_i\}$}{
					$F_{i} \leftarrow F_{i} \cup \{x\}$\label{line_a1}\;
					$\widehat{F} \leftarrow \widehat{F} \setminus \{x\}$\label{line_a2}\label{line_lemma_a_end}\;
				}
			}
		}
		\Else(\tcp*[h]{Use \Cref{lemma_rewrite}(b)}){\label{line_lemma_b_start}
			\ForEach(\tcp*[h]{Consider only literals in $x$}){$(\neg) \gamma \in x$}{\label{line_foreach_2}
					\If{
					$\mathbb{T}(\gamma) = \left\{ T_{i} \right\}$}
					{
					$\Gamma_i \leftarrow \Gamma_i \cup \{(\neg)\gamma\}$\;
					}
				}
				\ForEach{$i \in \{1, \dots s\}$} {\label{line_foreach_b_2}
					\If{$\Gamma_i \neq \varnothing$}{\label{lineb1}
						$\beta_{i} \leftarrow \beta_{i} \cup \{p_{x,i}: \bigvee_{(\neg)\gamma \in \Gamma_i}((\neg)\gamma)\}$\label{line_betai_pi}\;
						$\widehat{x} \leftarrow x \setminus \Gamma_i \cup \{p_{x,i} = \texttt{true}\}$\;
						$\widehat{F} \leftarrow \widehat{F} \cup \{\widehat{x}\}$\label{err_line_pi_true}\;
						$\widehat{F} \leftarrow \widehat{F} \setminus \{x\}$\label{line_lemma_b_end}\label{lineb2}\;
					}
				}
				}
				}
		\ForEach{$x \in \widehat{F}$}{\label{line_c1} 
			\ForEach(\tcp*[h]{Literals written as $A\theta B$ $(\theta \in \{=, >\})$}){$(\neg) A\theta B \in x$}{ 
				$\{T_i, T_j\} \leftarrow \mathbb{T}(A\theta B)$\;
				$A_i \leftarrow \mathbb{A}(T_i) \cap \{A,B\}$\;
				$A_j \leftarrow \mathbb{A}(T_j) \cap \{A,B\}$\;
			$\beta_{i} \leftarrow \beta_{i} \cup A_i$\;
			$\beta_{j} \leftarrow \beta_{j} \cup A_j$\;
	}
}
	\ForEach{$A \in \beta$}{\label{line_foreach_3_end}
		$i \leftarrow \{i | A \in \mathbb{A}(T_i)\}$\; 
		$\beta_i \leftarrow \beta_i \cup A$\;
	}\label{line_c2}
	\ForEach{$i \in \{1,\dots s$\}}{
		\If{$F_i = \emptyset$}{
			$F_i \leftarrow$ \{\{\texttt{true}\}\}\;
		}
		$\widehat{q_i} \leftarrow \pi_{\beta_i}\sigma_{{F_i}}(T_i)$\label{line_def_hatq}\;
		\If{$\beta_i = \emptyset$}{
			$\widehat{q_i} \leftarrow q^{\varnothing}$\label{line_def_hatq2}\;
		}
	}
	$\widehat{q} \leftarrow \pi_{\beta}\sigma_{\widehat{F}}(R_1 \Join \dots \Join R_s)$\label{line_def_q}\;
	\Return{$\widehat{q},\widehat{q_1}, \dots \widehat{q_s}$}
	\caption{Query Distribution}\label{algo_distributed_queries}
\end{algorithm}

\begin{theorem}\label{theoremOne}
	\Cref{algo_distributed_queries} solves the CS-Problem, for all queries in extended normal form.
\end{theorem}
\begin{proof}
	We can easily see, that \Cref{algo_distributed_queries} fulfills the syntactical requirements of \Cref{problemOne}.
	It remains to show that $q$ evaluated on $T_1, \dots T_s$ is equal to $\widehat{q}$ evaluated on $R_1, \dots R_s$, where $R_i$ is the evaluation of $\widehat{q_i}$ on $T_i$. 
	The proof is structured into substatements about the algorithm, utilizing \Cref{lemma_rewrite}.
	In the following, we will examine the query $Q$ at different intermediate states of the algorithm.
	We assume that $Q$ is constructed from the intermediate values
	of~$\beta, \widehat{F}, \beta_i, {F_i}$ as defined in \Cref{prop_Q_form}.
	At the end of the algorithm, $Q$ = $\widehat{q}$.

	\begin{statement}\label{prop_Q_form}
	$Q$ is from the form $\pi_\beta\sigma_{\widehat{F}} (\Join_{i=1}^{s}\pi_{\beta_i}\sigma_{{F_i}}(T_i))$, making \Cref{lemma_rewrite} generally applicable to it.
	\end{statement}
	This follows directly from \Cref{line_def_hatq,line_def_q},
	the construction of $Q$, and the fact that
	$R_i$ is the evaluation of $\pi_{\beta_i}\sigma_{F_i}\widehat{q_i}$ on $T_i$.

	\begin{statement}\label{prop_foreach_1}
	The foreach-loop in \Cref{line_foreach_1} preserves equivalence for $Q$.
	\end{statement}
  Let $Q_0$ be the intermediate query before the start of the foreach-loop
  and $Q_k$ the intermediate query after the $k$-th round.
  We need to show that for $k = 1, \dots n$, it holds that $Q_{k-1} \equiv Q_k$.
	Let $k \in \{1, \dots n\}$ and $x_k$ be the clause of the $k$-th round.
	If there is a $T_i \in \{T_1, \dots T_s\}$ with $\mathbb{T}(x_k) = \{T_i\}$, Lines~\ref{line_lemma_a_start}--\ref{line_lemma_a_end} rewrites $Q_{k-1}$ to $Q_k$ according to \Cref{lemma_rewrite}~(a).
	If $x_k$ can not be fully evaluated in one table the code in Lines~\ref{line_lemma_b_start}--\ref{line_lemma_b_end} repeatedly rewrites $Q_{k-1}$ using \Cref{lemma_rewrite}~(b).
	In \Cref{line_foreach_b_2} every $\Gamma_i$ only holds $\gamma$ with $\mathbb{T}(\gamma) = \{T_i\}$.
	As such the $\Gamma_i$ are subsets, as defined in \Cref{lemma_rewrite}~(b).
	The Lines~\ref{lineb1}--\ref{line_lemma_b_end} only execute the Lemma for each of the subsets~$\Gamma_i$, making $Q_k$ equivalent to $Q_{k-1}$.

\begin{statement}\label{statement_3}
	The code from \Cref{line_c1} to \Cref{line_c2} is preserves equivalence for $Q$.
\end{statement}
This property holds because at \Cref{line_c1}
  $\widehat{F}$ contains only literals $\gamma$
  with $|\mathbb{T}(\gamma)| = 2$,
  as literals with $|\mathbb{T}(\gamma)| =1$
  are removed in \Cref{line_a2,lineb2}.
  Consequently, in \Cref{line_foreach_3_end},
  every $\beta_i$ holds the attributes that $\widehat{F}$
  uses from table $T_i$ $(i=1, \dots s)$.
The foreach-loop starting at \Cref{line_foreach_3_end}
ensures that $\beta_i = \mathbb{A}(T_i, F, \beta) \cup P_i$
where~$\mathbb{A}(T_i, F, \beta)$ is as defined in \Cref{lemma_rewrite}~(c) and $P_i$ are values held $\beta_i$ before \Cref{line_c1}.
Since for projections $\pi_{A}\pi_{B}$ can be replaced by $\pi_{A \cup B}$ for every pair of sets $A,B$, \Cref{statement_3} follows by using \Cref{lemma_rewrite}~(c) for all $i \in \{1, \dots s\}$.

The theorem follows from \Cref{prop_Q_form,prop_foreach_1,statement_3}.
\end{proof}

\section{Analysis of Information Leakage}\label{analysis}
In this section, we quantify the information leakage
incurred by a central exchange when solving an instance of \Cref{problemOne}.
We show that \Cref{algo_distributed_queries} minimizes leakage,
regardless of the exchange's prior knowledge.

To measure leakage, we assess the probability of the data exchange
correctly reconstructing the full tables $T_1,\dots T_s$
based on $R_1 = \widehat{q_1}[T_1], \dots R_s = \widehat{q_s}[T_s]$.
We consider $T_i$ as a set of tuples over the cross-product of its attribute domains,
$\mathbb{D}(T_i) := \bigtimes_{A \in \mathbb{A}(T_i)} \mathbb{D}(A)$,
disregarding row order.
The exchange's assumptions about~$T_i$ are modeled by the probability measure~$p^i$,
sampling from all possible tables $\tilde{T} \in \mathbb{D}(T_i)^2$.
We consider only discrete $p^i$,
where $p^i(\tilde{T}) > 0$ holds for a finite number of $\tilde{T}$,
reflecting real-world scenarios where attribute domains are naturally finite.
A relational exchange updates its assumptions for~$T_i$ upon receiving a result~$R_i$.
The updated assumptions are modeled by $p^i_{R_i}$.
To determine the optimal solution for \Cref{problemOne},
we evaluate the deviation of $p^i_{R_i}$ for all $i \in \{1, \dots s\}$
from correctly guessing $T_i$,
where $R_i$ is the (implicit) output of a solution.

To quantify the difference between two probabilities $p$ and $q$,
we use the Kullback-Leibler Divergence~\cite{kullback_information_1951}.
\begin{definition}[Discrete Kullback-Leibler Divergence~\cite{chambert-loir_information_2022}]
	For two discrete probability measures $p$ and $q$ with sample set $X$,
	where $q(x) = 0$ implies $p(x) = 0$,
	the \emph{Kullback-Leibler Divergence} of $p$ and $q$ (with $0 \log(0) := 0$)
	is defined as:
	\[ D(p||q) = \sum_{x \in X} p(x) \log{\frac{p(x)}{q(x)}}. \]
\end{definition}
For our case, we calculate the divergence between~$p^i_{R_i}$ and~$p^i_{T_i}$,
where $p^i_{T_i}$ fully reveals $T_i$.
Specifically, $p^i_{T_i}(T_i) = 1$ and $p^i_{T_i}(\tilde{T}) = 0$
for all $\tilde{T} \in \mathbb{D}(T_i)^2 \setminus \{T_i\}$.
From
\begin{equation}\label{eq_KL-D}
	D(p^i_{T_i}||p^i_{R_i}) =
	\sum_{\tilde{T} \in \mathbb{D}(T_i)^2} p^i_{T_i}(\tilde{T}) \log\left(\frac{p^i_{T_i}(\tilde{T})}{p^i_{R_i}(\tilde{T})}\right) =
	\log\left( p^i_{R_i}(T_i)^{-1} \right)
\end{equation}
the divergence depends only on the value of $p^i_{R_i}$ at $T_i$.

\subsubsection{Calculating $p^i_{R_i}$}
Intuitively, $p^i_{R_i}$ should correspond to the initial probability~$p^i$
under the condition that $R_i$ is the result of the evaluation of $\widehat{q_i}$. 
However, using plain conditional probability~$p^i(X | R_i)$ is not feasible.
This is because~$R_i$ is not guaranteed to be an element of $\mathbb{D}(T_i)^2$,
as $\widehat{q_i}$ might utilize projections.
Nonetheless, the information gained from $R_i$ can be used to minimize
the set of possible candidates $\tilde{T} \in \mathbb{D}(T_i)$.
Since $R_i$ is queried from $T_i$, all entries in $R_i$ must have
exactly one compatible entry in $\tilde{T}$.
For compatibility, two conditions must be satisfied:
(1)~For all attributes of $R_i$ that do not result from propositional projections,
compatible entries must have the same values projected on these attributes in $\tilde{T}$.
(2)~For all columns originating from propositional projections,
the truth-value in that entry must match the evaluation of that entry in $\tilde{T}$.
Formally, we define the possible candidates for $T_i$ given $R_i$ as:
\begin{equation}\label{eq_def_m}
	\begin{split}
	C^i(R_i) = \{
		\tilde{T} \in \mathbb{D}(T_i)^2 &|
		\exists \text{ injective } m: R_i \hookrightarrow \tilde{T}, \\
		\forall x \in R_i:
																&\pi_{\mathbb{A}(R_i) \cap \mathbb{A}(T_i)}x = \pi_{\mathbb{A}(R_i) \cap \mathbb{A}(T_i)} m(x)
		\wedge \\
		\forall p: &\bigvee_{i}((\neg)\gamma_i) \in \mathbb{A}(R_i): \pi_{p}(x) = \bigvee_{i} (\neg)\gamma_i[m(x)]
	\}.
\end{split}
\end{equation}
This allows us to define $p^i_{R_i}$ as the probability of $p^i$ under the condition $C^i(R_i)$:
\begin{equation}\label{eq_pi_under_Ri}
	p^i_{R_i}(X) := p^i(X\, |\, C^i(R_i)) = \frac{p^i(X \cap C^i(R_i))}{p^i(C^i(R_i))}
\end{equation}
Note that if $R_i$ is part of a solution to \Cref{problemOne},
where $T_i$ is part of the input, then $T_i \in p^i(C^i(R_i))$.
As such, $p^i(C^i(R_i)) \neq 0$ if $p^i(T_i) > 0$.

\subsubsection{Solving the CS-Problem with Minimal Information Leakage}
We can now prove that \Cref{algo_distributed_queries}
is the optimal solution for \Cref{problemOne}.
\begin{theorem}\label{theo_optimum}
	Let $q = \pi_{\beta}\sigma_{F}(\Join T_i)$ be a query in RAPP in extended normal form over $T_1, \dots T_s$.
	Let $\widehat{q}, \widehat{q_1}, \dots \widehat{q_s}$ with $R_1, \dots R_s$ be the solution of \Cref{problemOne} given~$q$ produced by \Cref{algo_distributed_queries}.
	Let $\tilde{q}, \tilde{q_1}, \dots \tilde{q_s}$ with $\tilde{R_1}, \dots \tilde{R_s}$ be any other solution of \Cref{problemOne} given $q$.
	For any set of discrete probability measures~$p^1, \dots p^s$ over~$\mathbb{D}(T_1)^2, \dots \mathbb{D}(T_s)^2$ respectively with $p^i(T_i) > 0$ for any~$i \in \{1,\dots s\}$:
	\[
		D(p^i_{T_i}||p^i_{R_i}) \geq D(p^i_{T_i}||p^i_{\tilde{R_i}})
	\]
\end{theorem}
\begin{proof}
	Let $i \in \{1, \dots s\}$.
	With $T_i \in C^i(\tilde{R_i})$, \Cref{eq_KL-D,eq_pi_under_Ri},
	and the additive property of probability measures, the following holds:\\
	\resizebox{\textwidth}{!}{$%
		D(p^i_{T_i}||p^i_{R_i}) \geq D(p^i_{T_i}||p^i_{\tilde{R_i}})
		\Leftrightarrow
		p^i(C^i(R_i)) \geq p^i(C^i(\tilde{R_i}))
		\Leftrightarrow
		\sum_{\tilde{T} \in C^i(R_i)}p^i(\tilde{T}) \geq
		\sum_{\tilde{T} \in C^i(\tilde{R_i})}p^i(\tilde{T})
		.$}

		Because of this, it suffices to show that $C^i(R_i) \supseteq C^i(\tilde{R_i})$.
		We will show this by proving four substatements.

		\begin{statement}\label{stat_Attr}
			$\mathbb{A}(R_i) \cap \mathbb{A}(T_i) \subseteq \mathbb{A}(\tilde{R_i}) \cap \mathbb{A}(T_i)$
		\end{statement}
		Let $A \in \mathbb{A}(R_i) \cap \mathbb{A}(T_i)$.
		By definition of $R_i = \pi_{\beta_i}\sigma_{F_i}(T_i)$,
		  it follows that $A \in \beta_i$.
	  From the construction of $\beta_i$ in Lines~\ref{line_foreach_3_end}--\ref{line_c2}
	    of \Cref{algo_distributed_queries}, it holds that $A \in \beta$.
	  Since $q(\Join T_i) = \tilde{q}(\Join R_i) = \tilde{q}(\Join \tilde{q_i}(T_i))$,
	    it follows that $A \in \mathbb{A}(\tilde{R_i}) \cap \mathbb{A}(T_i)$.

		\begin{statement}\label{stat_proj}
			For every $x \in F$ with $\{T_i, T_j\} \subseteq \mathbb{T}(x), (j \neq i)$ 
			and every $e \in \tilde{R_i}$ there must exist a projection $\pi_{p}$ such
			that $\pi_{p}e = \texttt{true}$ iff $\bigvee_{(\neg)\gamma \in \Gamma_i}(\neg)\gamma(e) = \texttt{true}$,
			where $\Gamma_i = \{(\neg)\gamma \in x \,|\, \mathbb{T}(x) = \{T_i\}\}$.
			Furthermore, $\pi_{p \cup \mathbb{A}(\tilde{R_i})}\tilde{R_i}$ is also a solution to \Cref{problemOne}.
		\end{statement}
		This statement directly follows from the fact that $\tilde{R_i}$ is part of a solution to \Cref{problemOne} for $q$.
		Since $x$ depends on values from at least $T_i$ and $T_j$,
		$\tilde{R_i}$ must indicate, for each of its entries $e$,
		whether that entry sets $x$ to \texttt{true}.
		By definition, this is the case if $\bigvee_{(\neg)\gamma \in \Gamma_i}(\neg)\gamma(e)$ is \texttt{true}.

	  \begin{statement}\label{stat_r}
			For all $\tilde{T} \in C^i(\tilde{R_i})$, there exists $\beta$ and $F$
			such that $\pi_{\beta}\sigma_{F}(\tilde{T}) = \tilde{R_i}$.
	  \end{statement}
	    Let $\beta = (\mathbb{A}(\tilde{R_i}) \cap \mathbb{A}(\tilde{T}) ) \cup \{p: (\neg)\gamma | p: (\neg)\gamma \in \mathbb{A}(\tilde{R_i})\}$
	    and $F$ be defined as $F(e) = \texttt{true}$ iff $e \in Img(\tilde{m})$.
	  Here, $Img(\tilde{m})$ represents the image of the function
	    $m: \tilde{R_i} \hookrightarrow \tilde{T}$, as defined in \cref{eq_def_m}.

	\begin{statement}\label{stat_img}
		Let $M_i: R_i \hookrightarrow T_i$ and $\tilde{M_i}: \tilde{R_i} \hookrightarrow T_i$ be the functions for $T_i \in C^i(R_i) \cap C^i(\tilde{R_i})$ in \cref{eq_def_m}.
		Then, $Img(M_i) \subseteq Img(\tilde{M_i})$.
	\end{statement}
	We prove this statement by contradiction.
	Let $e \in Img(M_i) \setminus Img(\tilde{M_i})$.
	Let $\beta'$, $F'$, $\tilde{\beta_i}$, and $\tilde{F}$ be such that
	  $\pi_{\beta'}\sigma_{F'}(T_i) = R_i$
	  and $\pi_{\tilde{\beta_i}}\sigma_{\tilde{F}}(T_i) = \tilde{R_i}$
	  (as shown in \Cref{stat_r}).
	By construction, $F'(e) = \texttt{true}$ and $\tilde{F}(e) = \texttt{false}$.
	Since $R_i$ and $\tilde{R_i}$ are part of solutions of \Cref{problemOne},
	  it holds that $q(\Join T_i) = \widehat{q}(\Join R_i) = \tilde{q}(\Join \tilde{R_i})$.
	Therefore, $e$ cannot be a part of the evaluation of $q(\Join T_i)$.
	More precisely, for every $t \in q(\Join T_i)$ with $\pi_{\mathbb{A}(e)}t = e$: $F(t) = \texttt{false}$.
	Since this needs to hold independently of the other $T_j (j \neq i)$,
	there exists a clause $x \in F$ with $\mathbb{T}(x) = \{T_i\}$ and $x(e) = \texttt{false}$.
		Let $x$ be such a clause.
	We now look at the construction of $R_i$ in \Cref{algo_distributed_queries}.
	By \Cref{line_a1}, $x \in F_i$. Hence, $e \notin R_i$.
	This is a contradiction with $F'(e) = \texttt{true}$.

	Given the four statements,
	  we can show that $C^i(R_i) \subseteq C^i(\tilde{R_i})$.
	Let $\tilde{T} \in C^i(\tilde{R_i})$ with $\tilde{m}: \tilde{R_i} \hookrightarrow \tilde{T}$
	  and $M_i, \tilde{M_i}$ as described in \Cref{stat_img}.
	We need to show that there exists an $m: R_i \hookrightarrow \tilde{T}$
	  satisfying the properties in \Cref{eq_def_m}.
  We argue that $m: x \mapsto \tilde{m}(\tilde{M}_i^{-1}(M_i(x)))$ is such a function.
	The function is well-defined due to \Cref{stat_img}.
	Its domain is $\tilde{T}$, and it is injective as a concatenation of injective functions.
	The expression $\pi_{\mathbb{A}(R_i) \cap \mathbb{A}(T_i)} x = \pi_{\mathbb{A}(R_i) \cap \mathbb{A}(T_i)} m(x)$ is only a problem
	  when mapping $M_i(x)$ to $\tilde{M}_i^{-1}(M_i(x))$.
	However, since $\tilde{M}_i^{-1}$ satisfies the required property,
	  it holds that $\pi_{\mathbb{A}(\tilde{R_i}) \cap \mathbb{A}(T_i)} \tilde{M}_i^{-1}(M_i(x)) = \pi_{\mathbb{A}(\tilde{R_i}) \cap \mathbb{A}(T_i)} M_i(x)$,
	   and the property follows from \Cref{stat_Attr}.
		 From the construction of $R_i$ in \Cref{algo_distributed_queries},
		 we know that $R_i$ only has propositional projections if there exists a clause $x \in F$ with $\{T_i, T_j\} \subseteq \mathbb{T}(x), (j \neq i)$,
		 and all have the form $p_{x,i}: \bigvee_{(\neg)\gamma\in\Gamma_i}(\neg)\gamma$.
		 To fulfill the second property of~\Cref{eq_def_m},
		 we assume WLOG that $\tilde{R_i}$ has attributes $\tilde{p}_{x,i}$
		 for all $p_{x,i}$ which are true iff the disjunction of $p_{x,i}$
		 above is \texttt{true} (as described in \Cref{stat_proj}).
		 With this, $m$ satisfies the second property of~\Cref{eq_def_m}, because $\tilde{m}$ satisfies it.
\end{proof}

\section{Related Work}\label{discussion}
The concept of data trustees aligns with models used in distributed database systems,
where data is managed across multiple sources.
Unlike distributed databases, which often replicate data across nodes,
data trustees typically manage unique data from different sources.
An overview of distributed database principles
can be found in~\cite{xu_native_2024}.

To minimize information leakage, we push query execution
as close to the data sources as possible.
This approach is different to the late materialization technique
used in database management systems
to enhance performance~\cite{chernishev_comprehensive_2022}.
Despite these advantages, PrivTru currently does not support non-relational data formats,
which limits its applicability in certain cases.
For example, sharing medical data such as MRI images
from multiple sources poses challenges,
as discussed by Kolain and Malavi \cite{kolain_zukunft_2019}.
Extracting only the minimum required information from such detailed data
remains a significant challenge.

Beyond the privacy strategies implemented in our work,
systems can be evaluated using frameworks
such as the LINDUNN~\cite{cavoukian_privacy_2012}
or the privacy protection goals proposed in~\cite{hansen_protection_2015}.
We believe PrivTru fulfills their requirements as well.

%

\section{Conclusion}\label{conclusion}
In this paper, we introduced PrivTru, a privacy-by-design data trustee,
which facilitates data exchange from data sources to data receivers.
We contributed to the ongoing discussion on whether data trustees
should act as data stewards (storing data)
or as data exchanges (relaying information).
We argue that data exchanges are preferable,
as they comply with the \textsc{Separate} design strategy
while also adhering to the principles of
\textsc{Hide}, \textsc{Minimize}, and \textsc{Aggregate}.
PrivTru exemplifies an instantiation of data exchanges,
achieving \textsc{Minimize} by ensuring
that each data source provides only the minimum data necessary to answer a query.
We proved that this property holds regardless of the exchange's prior knowledge.
While we did not explore \textsc{Aggregate} in detail,
it can be achieved by allowing queries with aggregation functions
to be evaluated at the data sources,
as done with other operations in PrivTru.
%
%
In summary, our work establishes a foundation for designing data trustees
that balance data utility with robust privacy protection,
contributing to a more trustworthy data economy.

\section*{Acknowledgements}
This research has been funded
by the Federal Ministry of Education and Research of Germany
(BMBF, Project~16KISA038).\\
The final publication is available at Springer via \\\url{https://doi.org/10.1007/978-3-031-92882-6_8}

\bibliography{refs}

\begin{thebibliography}{10}
\providecommand{\url}[1]{\texttt{#1}}
\providecommand{\urlprefix}{URL }
\providecommand{\doi}[1]{https://doi.org/#1}

\bibitem{abiteboul_foundations_1995}
Abiteboul, S., Hull, R., Vianu, V.: Foundations of {Databases} (1995)

\bibitem{aline_blankertz_designing_2020}
{Aline Blankertz}: Designing data trusts. {Why} we need to test consumer data
  trusts now (2020),
  \url{https://www.interface-eu.org/storage/archive/files/designing_data_trusts_e.pdf},
  accessed: 2024-12-11

\bibitem{blankertz_what_2021}
Blankertz, A., Specht, L.: What regulation for data trusts should look like
  (2021),
  \url{https://www.interface-eu.org/storage/archive/files/regulation_for_data_trusts_0.pdf},
  accessed: 2024-12-11

\bibitem{cavoukian_privacy_2012}
Cavoukian, A.: Privacy by {Design} [{Leading} {Edge}]. IEEE Technology and
  Society Magazine  \textbf{31},  18--19 (2012). \doi{10.1109/MTS.2012.2225459}

\bibitem{chambert-loir_information_2022}
Chambert-Loir, A.: Information {Theory}: {Three} {Theorems} by {Claude}
  {Shannon}, {UNITEXT}, vol.~144. Cham (2022). \doi{10.1007/978-3-031-21561-2}

\bibitem{chernishev_comprehensive_2022}
Chernishev, G., Galaktionov, V., Grigorev, V., Klyuchikov, E., Smirnov, K.: A
  {Comprehensive} {Study} of {Late} {Materialization} {Strategies} for a
  {Disk}-{Based} {Column}-{Store}. In: DOLAP. pp. 21--30 (2022)

\bibitem{codd_relational_1970}
Codd, E.F.: A relational model of data for large shared data banks. Commun. ACM
   \textbf{13},  377--387 (1970). \doi{10.1145/362384.362685}

\bibitem{hansen_protection_2015}
Hansen, M., Jensen, M., Rost, M.: Protection {Goals} for {Privacy}
  {Engineering}. IEEE Security and Privacy Workshops  (2015).
  \doi{10.1109/SPW.2015.13}

\bibitem{hoepman_privacy_2014}
Hoepman, J.H.: Privacy {Design} {Strategies}. In: ICT Systems Security and
  Privacy Protection. p.~446 (2014). \doi{10.1007/978-3-642-55415-5_38}

\bibitem{kolain_zukunft_2019}
Kolain, M., Molavi, R.: Zukunft {Gesundheitsdaten} (2019),
  \url{https://www.bundesdruckerei-gmbh.de/files/dokumente/pdf/studie_zukunft-gesundheitsdaten.pdf},
  accessed: 2024-12-11

\bibitem{kullback_information_1951}
Kullback, S., Leibler, R.A.: On {Information} and {Sufficiency}. The Annals of
  Mathematical Statistics  \textbf{22},  79--86 (1951).
  \doi{10.1214/aoms/1177729694}

\bibitem{reiberg_datentreuhander_2023}
Reiberg, A., Appelt, D., Kraemer, P.: Data {Trusts}, {Data} {Intermediation}
  {Services} and {Gaia}-{X} (2023),
  \url{https://gaia-x-hub.de/wp-content/uploads/2023/11/GX-White-Paper-Data-Trusts.pdf},
  accessed: 2024-05-29

\bibitem{samarati_protecting_1998}
Samarati, P., Sweeney, L.: Protecting {Privacy} when {Disclosing}
  {Information}: k-{Anonymity} and {Its} {Enforcement} through {Generalization}
  and {Suppression}. Proceedings of the IEEE Symposium on Research in Security
  and Privacy  (1998)

\bibitem{xu_native_2024}
Xu, Q., Yang, C., Zhou, A.: Native {Distributed} {Databases}: {Problems},
  {Challenges} and {Opportunities}. Proceedings of the VLDB Endowment pp.
  4217--4220 (2024)

\end{thebibliography}
\bibliographystyle{splncs04}

\end{document}